\documentclass[]{article}
\usepackage{hpcs}
\usepackage{psfig}
\usepackage{graphicx}
\begin{document}

\title{
High Performance Commodity Networking in a
512-CPU Teraflop Beowulf Cluster for Computational Astrophysics
}

\author{John Dubinski, Robin Humble, Ue-Li Pen, Chris Loken and Peter Martin
\address[UdeS]{
Department of Astronomy and Astrophysics and Canadian Institute for Theoretical Astrophysics, 
dubinski@astro.utoronto.ca, rjh@cita.utoronto.ca, pen@cita.utoronto.ca, cloken@cita.utoronto.ca,
and pgmartin@cita.utoronto.ca}
}

\onecolumn
\begin{abstract}

We describe a new 512-CPU Beowulf cluster with Teraflop performance
dedicated to problems in computational astrophysics. 
The cluster incorporates a cubic network topology based 
on inexpensive commodity 24-port gigabit switches and point to point
connections through the second gigabit port on each Linux server.  
This configuration has network performance competitive with more 
expensive cluster configurations and is scaleable to much larger systems
using other network topologies.  Networking represents only about
9\% of our total system cost of USD\$561K.  The standard Top 500
HPL Linpack benchmark rating is 1.202 Teraflops on 512 CPUs so computing
costs by this measure are \$0.47/Megaflop. 
We also describe 4 different astrophysical applications using complex parallel algorithms
for studying large-scale structure formation, galaxy dynamics, magnetohydrodynamic
flows onto blackholes and planet formation currently running on the cluster
and achieving high parallel performance.  The MHD code achieved a sustained
speed of 2.2 teraflops in single precision or 44\% of the theoretical peak.

\end{abstract}

\maketitle

\section{Introduction}

The Olympic motto "Citius, altius, fortius" ("Swifter, 
higher, stronger") succinctly describes the direction of  
21st century parallel supercomputing - swifter processors, higher 
resolution and stronger fault-tolerant parallel algorithms. 
In computational astrophysics, the need for all of these qualities is perhaps 
greater than most supercomputing applications. While the 
bulk of the physics of the formation of the planets, stars, 
galaxies and the large-scale structure in the universe are 
now largely understood with the initial 
conditions well posed in some cases, the challenge of computing the 
formation of objects and structures in the universe is 
difficult.  The standard methods of computational fluid 
dynamics (including magnetohydrodynamics) and 
gravitational N-body simulation are stressed by the large 
dynamic range in density, pressure, and 
temperature that exist in nature.  Ever-finer 
computational meshes and greater numbers
particles in N-body simulations are needed to capture 
the physics of the formation of things in the universe correctly.

In recent years, the relative low cost of commodity 
servers running the Linux operating system and more 
importantly switches and network interface cards has led 
to the growth in performance and competitiveness of 
Beowulf clusters versus vector machines and symmetric multi-processors.  
The recent list of the Top 
500 supercomputers~\cite{top500} last year reveals 
that 55 out of the top 100 are clusters.
Most high performance Beowulfs now use low-latency,  proprietary
networking with costs several times that of current 
commodity gigabit networking. Large-scale 200+ port 
gigabit switches are a slightly cheaper option but still eat 
significantly into the cost of a cluster. Networking costs 
alone can dominate cluster costs.  

The spirit of building a Beowulf cluster is to get the highest
performance on a limited budget so the incorporation of expensive
networking solutions defeats this fundamental motivation.  
We describe here a 
cheaper networking configuration based on off-the-shelf 24-port 
commodity gigabit switches in a 256-node/512-CPU 
cluster.  We achieve a total cross-sectional bandwidth of 128 Gbit
between 256 nodes.
The innovation to achieving this performance is a
networking topology involving both a stack of commodity 
24-port gigabit SMC switches and the configuration of 
each Linux server as a network router through a second 
gigabit network port.  Our economical networking 
strategy competes directly with both large-scale 200+ port 
switches and proprietary low-latency networking at a fraction of the 
cost.  The high throughput provided by Linux routing capability
makes this performance possible.  
We ran the HPL LINPACK standard benchmark and achieved 1.202 Tflops 
placing it in the top 50 computers according to 
the Top 500 list of November 2002. 

We also describe the performance and results of 4 parallel applications 
for astrophysical problems that are currently running on the cluster
spanning problems in large-scale structure formation, galactic dynamics,
magnetohydrodynamic flows on blackholes and planet formation.

\section{Cluster Hardware Specifications}

In this section, we briefly describe the design and hardware 
specifications of the cluster.  The hardware was 
assembled and installed by Mynix Technologies, 
Montreal, PQ at the University of Toronto~\cite{mynix}.  
The installation of the Linux operation system was
done using the OSCAR~\cite{oscar} cluster package
by our group.

\subsection{Compute Nodes}
The CITA 
Beowulf Cluster dubbed McKenzie is comprised of 268 
dual Xeon 2.4 Ghz rack-mounted format Linux servers 
(536 processors in total) distributed over 7 racks.
All the nodes are 1U format with the 
exception of two 2U format head nodes.  The hardware 
specifications for each server are given in Table 1.  From 
the total of 268 nodes, 256 nodes are dedicated to parallel, 
message-passing supercomputing, 8 nodes for code 
development and smaller scale applications, 2 spare nodes 
also running to act as hot replacements in case of 
hardware failures on the compute nodes and 2 head nodes.  
The main head node ``bob" contains the home disk space 
as well as the root space images and Linux kernels for the 
compute slave nodes.  All slave nodes boot through the 
network from bob using PXElinux allowing easy kernel 
upgrades and maintenance for the cluster nodes.  A 
secondary master node ``doug" mirrors bob's root and 
home directories on a daily basis and acts a backup in 
case of a bob crash.  This allows quick recovery and 
continued operation if bob fails.  

\subsection{Switches and Cables}

The cluster is networked using 19 SMC Tiger 24-port 
configurable gigabit switches.  Seventeen switches are 
used for the main 256-node cluster network while the 
development cluster is connected to its own switch and 
the 19th switch is reserved as a spare and connected to the 
running spare nodes.  We defer the discussion of the 
network topology to the section below.

\subsection{Rack Configuration}
The nodes and switches are mounted on 7 44U high racks. 
The main 256 compute nodes are mounted 
on the first 3 and last 3 racks.  The central 
rack is reserved for the head nodes, development nodes 
and spares as well as the stack of switches.  The switches 
are mounted on the back of the racks to simplify the 
cabling.  

\begin{table}
\begin{tabular}{ll}
\hline
Nodes: &       256 main compute nodes + \\
&			\ \ 8 development nodes + \\
            &  2 head nodes + 2 spares = 268 total \\
Chassis:&      Chenbro RM11802 1U format \\
Motherboard:&  Intel Westville SE7500WV2  \\
CPUS:       &  Dual Xeon 2.4GHz processors  \\
            &  E7500 chipset/512KB L2 cache,  \\
            &  400 MHz system bus \\
            &  536 CPUs in total \\
Memory:     &  1 Gbyte DDR-200 RAM \\
            &  268 Gbytes total \\
Disk Drives:&  2X80 Gbyte Seagate IDE drives \\
Networking: &  Dual Intel Pro/1000XT gigabit \\
Switches:   &  19 SMC Tiger SMC8624T 24-port \\
            &  managed  gigabit switches 1U format \\
Storage:    & 3TB Arena RAID array \\
\hline
Node cost:  & USD\$513K \\
Network cost:&  USD\$48 \\
\hline
Total cost: & USD\$561K
\end{tabular}
\caption{Mckenzie Part List and Cost}
\end{table}

\subsection{Cost}

The cost for the compute nodes plus storage and incidentals was USD\$513K.
The cost for the gigabit switches plus cabling was USD\$48K representing only
9\% of the total cluster cost of \$561K.

\section{Networking}

When considering options for networking the cluster, 
we decided against proprietary low-latency 
networking and large-scale 200+ port switches after 
realizing that they would consume a significant 
fraction of our modest budget.  The prices of 24-port gigabit 
switches this past year had dropped significantly
so we considered networking schemes that could take advantage 
of this inexpensive hardware.

Fat-tree networks that link switches together through a 
hierarchy are a straightforward configuration that provide 
connectivity but quite low cross-sectional bandwith.  
Each node comes with a second gigabit port, however, 
and could be exploited in some way using point to 
point connections with other machines.  With this in mind 
we came up with the following scheme.  Switches (or 
pairs of trunked switches) can be thought of as separate 
tightly coupled networks of up to a few dozen nodes.  These 
nodes can be assigned to vertices in some more complex 
network topology.  How can one connect these vertices to 
form a high-bandwidth global network linking all the 
compute nodes?   After loading a switch with compute 
nodes there are few ports left to link to other vertices.  
However, each compute node provides a single 
connection through its second port that can in principle be 
connected to a compute node on an adjacent network vertex.
If each linux node has routing enabled,
network traffic can be relayed through the second 
port to other nodes.  This was the strategy we ultimately
employed in setting up a high-bandwidth global network.

\subsection{The Fat-Tree Maintenance Network}

Before setting up our high performance global network, 
we first set up a simple fat-tree that we treat as a robust 
but lower-performance maintenance network.  This 
network was used to install the cluster originally but also
acts as a bootstrap for configuring a high-performance network.  
In our final configuration, we trunk pairs of switches together making 
network vertices containing 48 ports with a total of 8 
vertices using 16 switches.  We connect 32 compute 
nodes to each vertex.  For 6 vertices, we run a 4-port 
trunk to a 17th master switch filling it to capacity.  The 
fat-tree networking is completed by connecting the last 
two vertices with 4-port trunks to 2 vertices trunked to the 
master switch.  The head nodes bob and doug, plus the 
development and spare nodes are all plugged into this 
network as well allowing direct communication between 
all available nodes.

This fat-tree network configuration is robust and provides 
connectivity to all the 256 nodes and is adequate for 
maintenance, installation and embarassingly parallel 
applications.  However, the cross-sectional bandwidth is 
not at all ideal for heavy duty MPI applications.  We now 
describe how we use this fat-tree network to bootstrap to 
our  high-bandwidth network which runs predominantly 
through point to point connections between the second 
network ports of all machines.

\subsection{The Cross-Diagonally Connected Cube Network}

\begin{figure*}
\centerline{\psfig{file=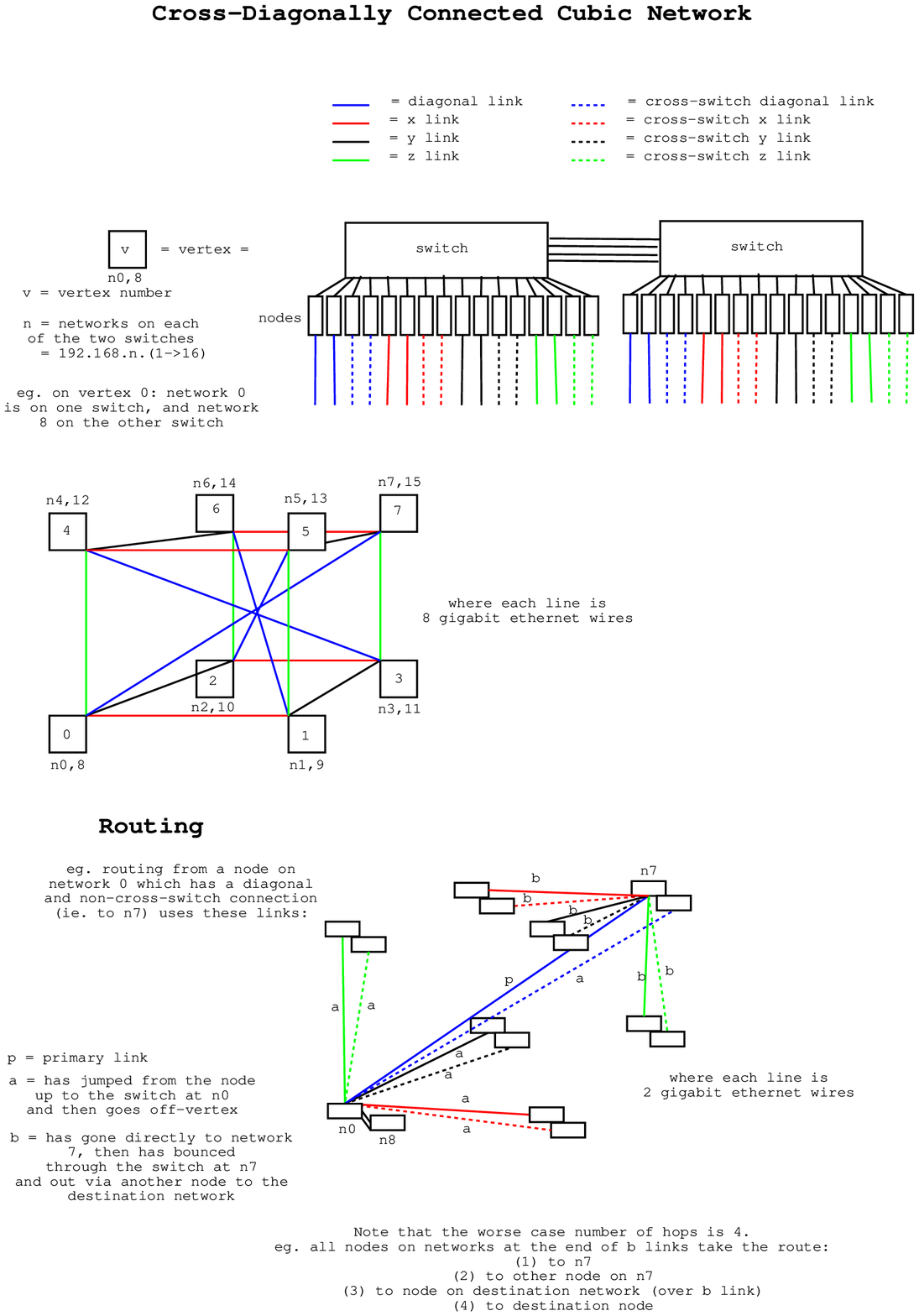,width=6.5in}}
\caption{
Cross-Diagonally Connected Cubic Networking Topology, Cabling and Routing Scheme
}
\label{cdcc}
\end{figure*}

The network vertices described above can be thought of 
as independent networks that we can connect to each 
other according to some topology.  The use of a master 
switch to build a fat tree is one such topology but it has 
quite low cross-sectional bandwidth.  How can we use the 
second port on each compute node to connect these 
vertices to increase the network bandwidth and minimize 
network hops to reduce latency? There are many 
possibilities but we finally settled on a cubic network 
topology (Figure~\ref{cdcc}).

In our chosen topology, the 8 network vertices are 
arranged on the corners of a cube.  Communication 
between vertices can occur along the edges of the cube 
but we also connect opposite corners through diagonals 
that cross through the cube centroid.  We call this 
topology a Cross-Diagonal Connected Cube (CDCC).
Each vertex then has 4 outgoing lines 
of communication to its adjacent corners and opposite 
along the diagonal.

The CDCC topology requires hard-wired routing tables 
between nodes which we constructed using a variety
of scripts.  We have 32 gigabit 
lines on each vertex provided by the second network port 
on each compute node and 4 vertex connections.  We 
therefore assign 8-gigabit lines to each vertex connection.  
These lines connect 8 compute nodes on one vertex to 
another 8 on a neighbouring vertex through direct port to 
port connections with a straight-thru cable.  In this way, 
there is an 8-gigabit pipeline connecting each vertex 
which must be shared between 32 compute nodes.  Also, 
the 4-port trunks connecting pairs of switches are 4-
gigabit pipelines relaying traffic for 16 compute nodes per 
switch.

In this way, the CDCC topology with 256 nodes 
approximately represents a fully-switched network 
running at 250 Mbit, although it is slightly better 
since nodes on the same switch are fully-switched at 1 
Gbit.  The full-duplex cross-sectional bandwidth is 128 Gbit
for 256 nodes.
When constructing the routing tables, the maximum 
number of network hops to get from one node to another 
is 4 so latency is greater than a fully-switched system 
but we find in practice for our applications and
standard benchmarks this is not a large hindrance.
The performance of this configuration is about 
half the 500 Mbit performance that is expected for large 
gigabit switches that typically connect 16 ports to an 8-
gigabit backplane internally.  Latency is also about 4 
times greater.  However, the total cost for the 17 required 
24-port switches plus cabling is much less than the
cost of a single 256-port capable switch.  We show below that our cluster 
provides competitive benchmark speeds for similar 
clusters with better networking and performs well for our 
applications.

\subsection{Cabling}

The task of cabling the cluster also proved to be 
formidable but we came up with some ways to minimize 
confusion and tangling.  Connections from the first 
ethernet port to the switches is straightforward if the 
switches are centralized to one location at the centre of 
the rack array in a stack.  Network cables are run over and 
under the cluster in small bundles and colour coded by 
rack number and then plugged into their appropriate ports 
in the central stack of switches.   Although tedious, this 
stage of the cabling took a small team of people two days to 
complete.  We point out that the same cabling effort is 
required even for more expensive multi-port switches so 
there is no significant overhead in manual labour.

In the second stage, it was necessary to wire the CDCC 
network using the point to point connections through the 
second network port on each node. Wiring these 
connections could potentially lead to a tangled mess of 
cables if the nodes are ordered sequentially according to 
their vertex numbers. A simple solution to this wiring 
problem is to re-order the nodes on the racks in logically
connected pairs so that they
are also physically connected to one another.   
In this way, only a 
short 1 foot length of cable is required for each contiguous
pair and wiring can be 
done quickly.  The logic of the connectivity of the CDCC 
network was used to create the correct ordered list of 
nodes on the rack for this arrangement as well as a wiring 
diagram for the connecting cables to the right switches in 
the fat-tree diagram.  The only extra level of complexity 
for doing things this way in comparison to 200+ port 
monster switches is the node ordering and the need to 
make sure each cable is connected to the correct port in 
the switch stack.  In practice, we have re-wired the 
network a couple of times to test the performance of 
different topologies and it generally only takes a few 
hours to do.

\subsection{Routing}

Another essential (and complex) part of getting the CDCC 
network to perform was the configuration of the routing 
tables for each Linux server.  Network packets relayed 
through each server on this server have to know where to 
go according to our defined network topology.

A useful part of our design is that we have a back-up 
maintenance network (the fat-tree) that we can always use 
to gain access to each node in case of routing bugs or 
broken links in the CDCC network.    This allowed us to 
experiment extensively with the scripts for generating the 
routing tables of  the CDCC and led to an optimized 
system.  Our strategy is to generate purely static routing 
tables on each server with a simple script at startup.  Each 
of these scripts contains routing commands that establish 
about 50 static routes to individual nodes and sub-networks 
(vertices) in the CDCC network.  Each node has
its own unique routing table.  We find this 
system works very well in practice and can be taken up 
and down quickly if necessary.

Our arrangement is also very fault-tolerant in case of node 
failures or broken links.  If one or more nodes fails in the 
system,  it will create holes in the CDCC network but as a 
contingency we can route around the broken nodes simply 
by using the fat-tree network.  We have written a simple 
network repair script that pings all neighbours on the 
CDCC - if a node fails to answer we attempt to establish a 
direct route to it through the fat-tree - if that fails the node 
is assumed to be dead.  In this way, we can continue to 
run jobs on the cluster even when main compute nodes go 
down.  Either the hot spares, or development nodes can 
fill in while we replace the failed nodes.  Since our networking
is similarly fragmented into many switches, the hardware
failure of one switch will only take out those nodes
connected to it so most of the cluster can still run while
the broken switch is being replaced.

Figure~\ref{cdcc} shows some final details of our routing scheme 
that are worth pointing out.  It turned out that the network 
performance of the trunks was not as good as we 
expected.  We therefore modified our routing scheme to 
avoid using the trunks as much as possible.  Network 
traffic from one node to another on different vertices used 
cross links that avoid the trunks.  The only network traffic 
going through the trunks is between the nodes on bonded 
pairs of switches.

\subsection{HPL Benchmark}

As an initial test of performance,
we ran the High-performance Linpack Benchmark 
(HPL), a portable implementation for 
distributed-memory computers 
to measure the speed of our cluster 
and compare with others on the Top 500 list~\cite{top500}.  This 
code uses the Message Passing Interface (MPI) libraries. 
The actual benchmark involves the inversion of the 
largest matrix that can be stored on your cluster.  In our 
case, the matrix size was $160000 \times 160000$ elements. We 
compiled the code using the Intel C compiler icc version 
7.0 and linked to the Intel MKL math libraries and 
Kazushige Goto's optimized BLAS libraries~\cite{goto}.   We 
used the LAM version 6.6b1 as the MPI library.  
We achieved a sustained speed of 1.202 Teraflops about 
48\% the theoretical peak speed of 2.46 Teraflops for 256 Xeon 2.4 GHz
chips.  This ratio of sustained to theoretical peak speed is comparable to many of the machines 
quoted on the November 2002 Top 500 list including 
similar sized clusters running with proprietary networking 
such as Myrinet.  Our ranking on the list would currently 
be number 34. Of course, it remains to be seen how we will 
rank on the next list released in a few months since these 
trends tend to evolve very quickly!

\begin{figure*}
\centerline{\psfig{file=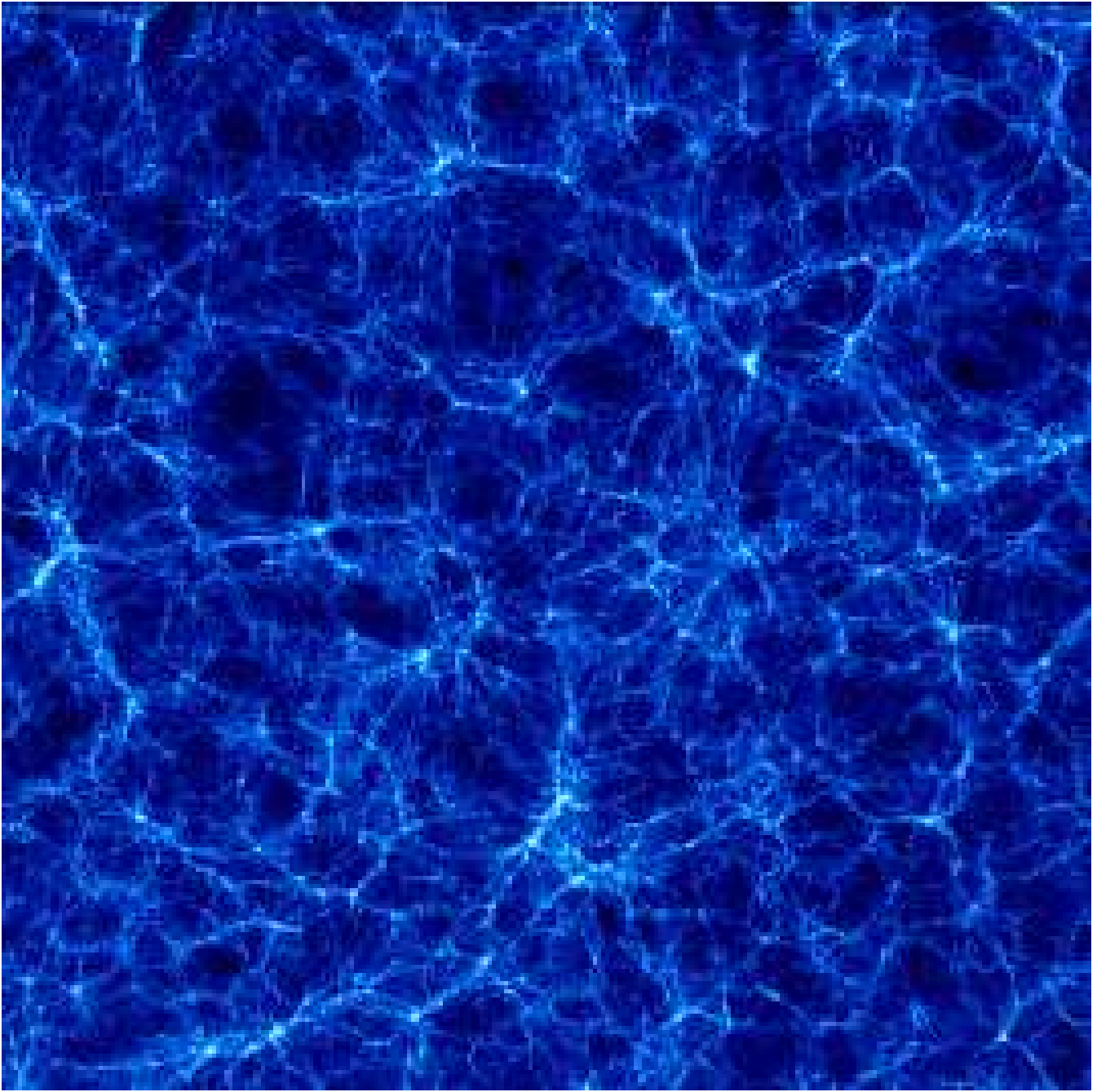,width=7.0in}}
\caption{A snapshot from a GOTPM N-body simulation of dark matter cosmology with matter density, $\Omega=0.3$
and dark energy, $\Lambda=0.7$ within a 200 $h^{-1}$ Mpc box.  Filamentary structure and clustering
are very well-resolved using a $1536^3$ mesh with $768^3$ particles.
}
\label{fig-cosmo}
\end{figure*}

\section{Astrophysical Applications}

The Beowulf cluster Mckenzie was constructed for problems in 
computational astrophysics.  We describe here several applications
that are being used to investigate areas on many astronomical
scales from planets to the entire universe.  We also highlight
the performance of the various codes we have started to use.

\subsection{Cosmology}

Cosmological N-body simulation is the main numerical tool used to study the
formation of structure in the universe.  Dubinski et al.~\cite{dub03}
have developed a new parallel algorithm that combines the standard techniques
of particle-mesh methods (PM)~\cite{hockney81} with the Barnes-Hut (BH) oct-tree 
algorithm~\cite{barnes86}.   The PM method is used to measure the long-range
gravitational forces from particles within a periodic cube representing a
typical volume in the universe.  These forces are refined on the sub-mesh scale
using the BH algorithm by building a grid of oct-trees within the simulation domain.
The new code is called GOTPM for Grid-(of)-Oct-Trees-Particle-Mesh.
The code uses slab domain decomposition to distribute the particles among
the CPUs.  During the PM phase, Poisson's equation for the gravitational potential
is solved using Fourier methods.  We use the MPI implementation of FFTW~\cite{frigo98}
to do the FFT's during this phase which assumes slabs of equal width.  
A grid of BH oct-trees is then constructed within
each slab and used to refine the forces at short range.  Force accuracies are typically
0.5\% using this method.  To ensure load balance, the slab widths are allowed to vary
during the tree phase with the width determined by the amount of computational work in the
previous timestep.  A fair amount of communication is required to move particle data at each
step since they must move back and forth from an equal width slab decomposition for the
PM phase and variable width slabs for tree phase.  Despite this high demand for bandwidth,
communication overhead is about 25\% even with our inexpensive networking configuration.

Figure~\ref{fig-cosmo} shows one snapshot from a cosmological simulation using a standard
model ($\Omega = 0.3$ and $\Lambda=0.7$) in a 200 $h^{-1}$ Mpc (650 million light years) box.
This simulation used a $1536^3$ mesh to solve Poisson's equation using Fourier methods in
single precision.  A total of $N=768^3$ (453M) particles are used and simulated for 5000 timesteps.
An animation of a fly-through of this dark matter  simulation as it structure develops is available at
our website~\cite{animations}.
In a 256 processor run, the wallclock time per timestep grows from 80 seconds at the beginning of the simulation to about
200 seconds once clustering develops with CPU time being dominated by the use of trees.  
The spatial resolution of these simulations is set by the gravitational softening radius 
used to shut down the $1/r^2$ Newtonian force at small separations.  For this simulation, we set 
the softening radius to 1 kpc (3000 light years) so we achieve a spatial resolution about 65 times
greater than what could be achieved using a PM method by itself.
\subsection{Galaxy Dynamics}

We have also applied a pure parallelized N-body treecode~\cite{dub96} to study galaxy interactions.
The main mode by which spiral galaxies are transformed into ellipticals is through merging.
When galaxies interact, the ordered energy of rotation and bulk motion of the two spiral 
galaxies is redistributed in the form of randomly oriented orbits.  This process 
only requires a few dynamical times and the final structure of the merger
remnant resembles and elliptical galaxy.

\begin{figure*}
\centerline{\psfig{file=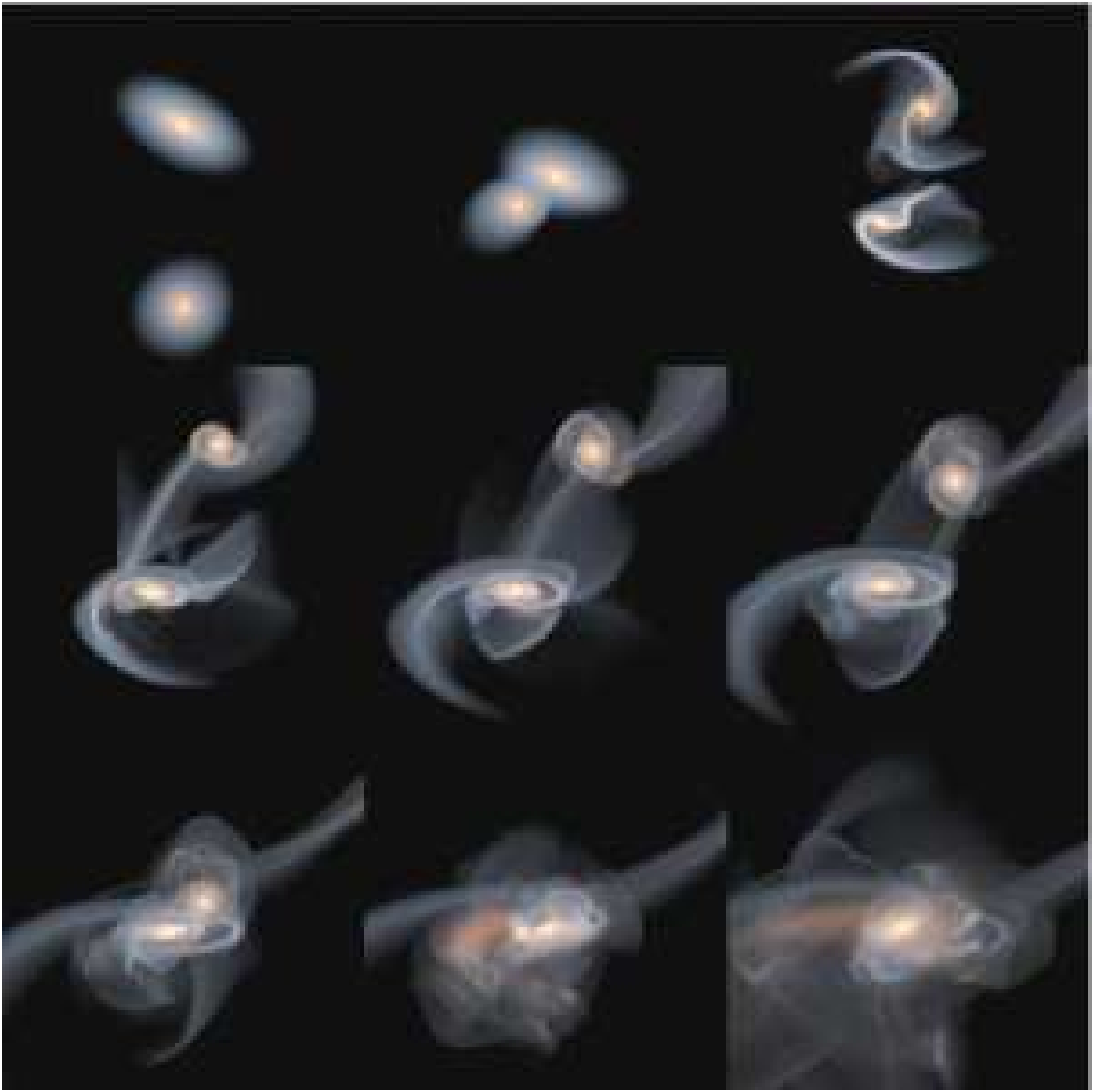,width=7.0in}}
\caption{
A simulation of the collision of two spiral galaxies modelled to represent the Milky Way and 
Andromeda galaxies.  Strong gravitational tidal forces excite open spiral structure
and throw off long tidal tails before the two galaxies merge into an elliptical galaxy.
}
\label{fig-gal}
\end{figure*}

To study this process at high resolution, we built two models of spiral galaxies composed
of a disk, a central bulge and surrounding dark halo using Kuijken \& Dubinski's method~\cite{kd95}.
To make the calculation more interesting, we used initial conditions chosen to represent a
future configuration of the Milky Way and the Andromeda Galaxy which will likely merge in about
3 billion years.  Each model is composed of 128M particles representing stars and 25.6 million particles
representing the surrounding dark matter halos for a total of 307.2M particles.  
The two galaxies are on a collision course and
we computed the trajectories for all particles for 5300 timesteps representing a physical 
timescale of 2.3 billion years.  A sequence of 9 images showing the evolution of the system
over this timescale is shown in Figure~\ref{fig-gal}.  An animation showing the galaxy collision is available 
here~\cite{animations}.

The code used for these calculations is PARTREE~\cite{dub96}, a parallelized treecode based on
Salmon's~\cite{salmon90} original algorithm and ideas of locally essential trees.  The code has 
recently been modified to incorporate asynchronous message-passing in the construction of the
locally essential trees to minimize communication overhead which becomes dominant for large
numbers of CPUs.  With current modifications, communication overhead only used 10\% of the wallclock
time of 180 seconds per step in the 307M particle simulation.  The code should scale
well to thousands of processors in its current form assuming the problem size grows as well.
A several thousand CPU machine with 2GB/CPU is now large enough to follow the trajectory of about 10 billion
particles, the same number of stars in some galaxies and only a factor of 40 away from the number of
stars in the Milky Way.  Galactic dynamics simulations will reach their ultimate resolution in terms
of the number particles representing stars in less than a decade.

\subsection{Blackhole Accretion}

\begin{figure*}
\centerline{\psfig{file=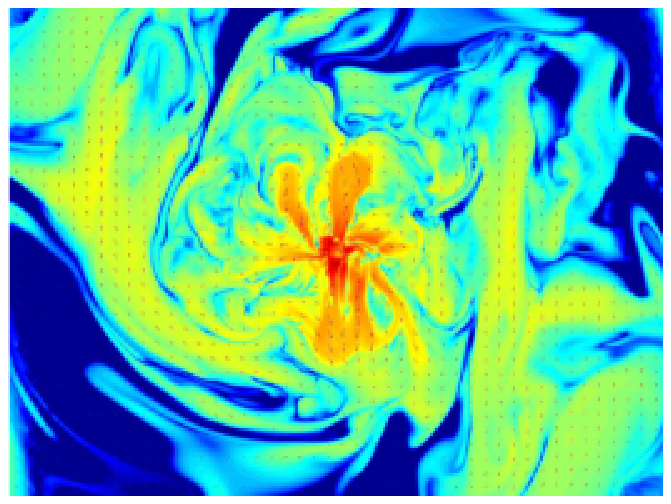,angle=90,width=5.0in}}
\caption{
Magnetic field structure: midplane ($x-y$) slice. Magnetic
pressure is shown along with projected magnetic field vectors.
This simulation describes the plasma flowing 
onto the supermassive black hole at the galactic center.  
The high resolution simulations are indicating for the first time that
the flow becomes clogged due the presence of magnetic fields.
}
\label{fig:mag}
\end{figure*}

We have undertaken MHD simulations with 1400$^3$ zones arrayed in a
uniform Cartesian grid, the largest MHD simulations to date (Figure \ref{fig:mag}).
At this resolution, each
full dimensional sweep corresponding to two timesteps took 40 seconds.
A series of optimizations allows the code to exploit the hardware's
vector units, hyperthreading OpenMP parallelism, and uses the Message
Passing Interface (MPI) to communicate between nodes.  The code~\cite{pen03a} 
is based on a 2nd order accurate in space
and time  high resolution Total-Variation-Diminishing (TVD) algorithm;
it explicitly conserves the sum of kinetic, thermal and magnetic energy;
hence magnetic dissipation (at the grid scale) heats gas directly.
No explicit resistivity or viscosity is added, and reconnection and
shocks occur through the solution of the flux conservation laws and the
TVD constraints.  Magnetic flux is conserved to machine precision by
storing fluxes perpendicular to each cell face.

Each time step requires about 20 seconds, 
corresponding to 2.2 Tflop sustained in single precision.  The operation
count was obtained by adding the number of operations in the source code
by hand.  This is about 44\% of theoretical peak speed achievable with the
using the SSE2 vector units.  The main parallelization overhead is the need for 16
buffer zones between computational domains.  This increases the floating
point count by 65\%.  Fatter nodes with more processors and more memory
would decrease the surface to volume ratio.  The domain decomposition
was matched to the physical three dimensional connectivity of the network.
Most traffic does not require any routing, and up both gigabit ports
per node can be effectively exploited.  All message passing latency
is hidden by staggering the computation phase: the processes do not
need to wait for data, and communication is asynchronous.


\begin{figure*}
\centerline{\psfig{file=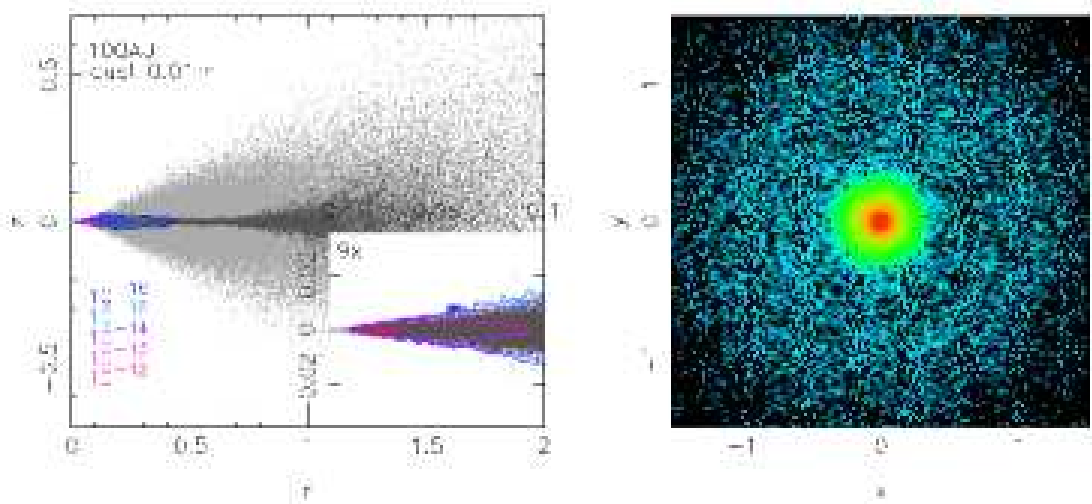,angle=270,width=150mm}}
\caption{Dust distribution after $10^4$ years in a protoplanetary disk
for dust grains of $1$cm radius. Axes are scaled in units of $100$AU.
The left hand r-z panel plots the dust distribution (dark grey)
over the gas disk (light grey) with contours of 3D dust density in g/cm$^3$.
The inset shows the closest $10$AU to the central protostar.
The right hand panel is a surface density map of the dust disk,
where blue is $\approx 10^{-3}$ g/cm$^2$ and red is $\approx 10$ g/cm$^2$.}
\label{fig-contour-1}
\end{figure*}

\subsection{Dusty disks and the Formation of planets}

The earliest stages of the planet formation process, in which micron sized particles
of dust somehow come together to form meter and kilometer sized objects and eventually planets,
is very poorly understood.
Humble, Maddison and Murray~\cite{rjh:03} have developed a new parallel  two-phase (dust
and  gas) smoothed particle hydrodynamics~\cite{jjm:97} and self-gravity code,
so that for the first time we can model the non-linear evolution of 
two coupled fluids in a 3D protoplanetary dusty disk.

A typical calculation uses 125k gas and
125k dust particles and resolves the motion of the entire disk from $2$ to $200$AU.
We investigate how dust evolves in a so-called minimum mass solar nebula in which
the total mass of the dust is $1$\% of the gas phase, which in turn is $1$\%
the mass of the protostar.
Figure~\ref{fig-contour-1} shows the distribution of $1$cm grains after $10^4$ years.

Our simulations clearly capture the expected settling and density enhancement of
the dust in the midplane of the disk where grain growth is most likely to occur,
and where gravitational instabilities may emerge.
The simulations also reveal the new result that the pressure supported differential
rotation between gas and dust is maintained only in a relatively narrow region in radius.
Rapid evolution of the dust occurs mostly in this narrow band, the location of which
is a strong function of the dust grain size. This results in
large variations of structure and dynamics between dusty disks of different
characteristic grain radii.

The code is based on the hashed oct tree algorithms of Salmon and Warren~\cite{sw},
is written in C and MPI and uses HDF5~\cite{hdf5} for i/o.
The code has evolved to be a flexible parallel tree framework, being tunable
for minimum bandwidth or maximum speed, and incorporating asynchronous MPI
latency hiding
techniques using multiple simultaneous tree traversals from each processor.
The hash table allows O($1$) random access to all
data in the tree and acts as a serialisation point for remote data retrieval,
whilst more traditional pointers are used for rapid traversal of all
locally cached tree data.

Most calculations performed to date utilise around $32$ processors with $90$\%
parallel efficiency. This number of processors is currently adequate
as this is a new field and there are large regions of parameter space to explore.
A relatively inexpensive machine like the McKenzie cluster is vital so that many runs
with different configurations and disk physics can be explored simultaneously
in a reasonable time.
We are working on further scalability and speed enhancements of the code. 

\section{Conclusions}

We have built Mckenzie, 512-CPU Beowulf cluster that incorporates commodity gigabit
networking in a new topology that permits 128 Gbit cross-sectional bandwidth.
Networking costs represent 9\% the total cost of the cluster of \$561K.  The machine
has been benchmarked with the Top 500 list HPL code and achieved a sustained
speed of $1.202$ Teraflops or 48\% of the theoretical peak speed making
it competitive with clusters with more expensive networking solutions.
By this measure, our computing costs are just \$0.47/Megaflop.
We note that the idea of using a Linux node as a router through a secondary
network or even tertiary network is a general concept that can be applied 
to different network topologies with larger numbers of nodes.  
The main limitation of this method is not so much the bandwidth but the
extra latency inherent to commodity switches and perhaps multiple hops
in the chosen topology.  Nevertheless, complex parallel algorithms with
high bandwidth requirements can still run on systems like this based on our
experience.

We have also described 4 astrophysical parallel applications running on the 
cluster that span the scale of the universe from solar system scale to 
the size of the universe.  We have been able to run some of the largest simulations
to date in cosmology, galaxy dynamics and MHD accretion flows using MPI parallel
codes.  N-body simulations of hundreds of millions of particles in cosmology
and galaxy dynamics can be done routinely.
The MHD code achieved a sustained speed of 2.2 teraflops in single
precision or 44\% of the theoretical peak speed in $1400^3$ zone calculation.  
Our low-cost networking 
also justifies using the machine in a parameter
survey mode on smaller sub-units of the cluster while such strategies might
be considered a waste of expensive resources on a large-scale SMP.
The ability to run on 512 processors has allowed us to tune and optimize
our codes to hide latency for large processor numbers and paves the way for
porting codes to clusters of thousands of processors.

\section*{Acknowledgments}

We acknowledge the Natural Sciences and Engineering Research Council of Canada
(NSERC) and the Canadian Foundation for Innovation (CFI) for funding
this project.

\bibliographystyle{unsrt}

\end{document}